\documentclass[twocolumn,nofootinbib,superscriptaddress]{revtex4}

\usepackage[english]{babel}
\usepackage{amsmath,amssymb}
\usepackage[dvips]{graphicx}

\newcommand{\CL}   {C.L.}
\newcommand{\dof}  {d.o.f.}
\newcommand{\EtAl} {{\it et al.\/}}
\newcommand{\eVq}  {\ensuremath{\text{eV}^2}}
\newcommand{\Dcq}  {\ensuremath{\Delta\chi^2}}
\newcommand{\Dms}  {\ensuremath{\Delta m^2}}

\newcommand{\AddrAHEP}{%
  AHEP Group, Instituto de F\'{\i}sica Corpuscular --
  C.S.I.C./Universitat de Val{\`e}ncia \\
  Edificio de Institutos de Paterna, Apartado 22085,
  E--46071 Val{\`e}ncia, Spain}

\newcommand{\AddrTUM}{%
  Theoretische Physik, Physik Department, 
  Technische Universit{\"a}t M{\"u}nchen,
  James--Franck--Strasse, D--85748 Garching, Germany}

\begin{document}

\preprint{IFIC/02-60}
\preprint{TUM-HEP-497/02}

\title{Combining first KamLAND results with solar neutrino data}

\author{M.~Maltoni}
\email{maltoni@ific.uv.es}
\affiliation{\AddrAHEP}

\author{T.~Schwetz}
\email{schwetz@ph.tum.de}
\affiliation{\AddrTUM}

\author{J.~W.~F.~Valle}
\email{valle@ific.uv.es}
\affiliation{\AddrAHEP}

\vspace{5mm}

\begin{abstract}
    We consider the impact of the recent KamLAND data on neutrino
    oscillations, the first terrestrial neutrino experiment that can
    probe the solar neutrino anomaly. By combining the first 145.1
    days of KamLAND data with the full sample of latest solar neutrino
    data we find an enhanced rejection against non-LMA oscillations,
    allowed only at more than $4\sigma$ with respect to
    LMA. Furthermore, the new data have a strong impact in narrowing
    down the allowed range of \Dms\ inside the LMA region.  In
    contrast, our global analysis indicates that the new data have
    little impact on the location of the best fit point.  In
    particular the solar neutrino mixing remains significantly
    non-maximal ($3\sigma$).
\end{abstract}

\maketitle

\section{Introduction}

In a recent paper the first results of the KamLAND collaboration
became public~\cite{kamlandPRL}. These data contain precious
information on the neutrino oscillation hypothesis which has been
advocated to account for a number of neutrino experiments involving
solar and atmospheric neutrinos and which indicate that neutrinos are
massive and that neutrino flavor mixing
occurs~\cite{Fukuda:2001nj,Fukuda:2002pe,Fukuda:1998fd,Fukuda:1998rq,Cleveland:nv,Davis:jw,Abdurashitov:1999zd,Hampel:1998xg,Altmann:2000ft,Cattadori:rd,sno,Fukuda:1998mi}.
The KamLAND experiment is a reactor neutrino experiment with its
detector located at the Kamiokande site. Most of the $\bar{\nu}_e$
flux incident at KamLAND comes from plants at distances of $80-350$ km
from the detector, making the average baseline of about 180
kilometers, long enough to provide a sensitive probe of the LMA
solution of the solar neutrino
problem~\cite{MSW,Bahcall:1999ed,Gonzalez-Garcia:1999aj}.
The KamLAND collaboration has for the first time measured the
disappearance of neutrinos traveling to a detector from a power
reactor.
They observe a strong evidence for the disappearance of neutrinos
during their flight over such distances, giving the first terrestrial
confirmation of the solar neutrino anomaly and also establishing the
oscillation hypothesis with man-produced neutrinos.
Moreover the parameters that describe this disappearance in terms of
the oscillations of the electron neutrino type to another, are
consistent with latest pre-KamLAND
determinations~\cite{Maltoni:2002ni,Bahcall:2002hv,Bandyopadhyay:2002xj,Barger:2002iv,deHolanda:2002pp,Creminelli:2001ij,Fogli:2002pt}
of solar neutrino parameters.

In this note we analyze the implications of these fundamental results
by combining the KamLAND data with data from solar neutrino
experiments. We will assume CPT conservation and for simplicity we
consider a two-flavor massive neutrino oscillation framework.  In
Sec.~\ref{sec:analysis} we analyze the impact of the KamLAND results
by including the full information on the spectral distribution of the
observed events.  Subsequently, in Sec.~\ref{sec:results} we perform a
global fit that combines the full KamLAND and Chooz reactor data
sample~\cite{Apollonio:1999ae} with the full solar neutrino data as
included in Ref.~\cite{Maltoni:2002ni}. In Sec.~\ref{sec:stability} we
check the stability of the results with respect to changes in the
statistical analysis, and we summarize in Sec.~\ref{sec:conclusions}.

\section{Simulation and analysis of KamLAND data}
\label{sec:analysis}

In KamLAND the target for the $\bar{\nu}_e$ flux consists of a
spherical transparent balloon filled with 1000 tons of non-doped
liquid scintillator. The anti-neutrinos are detected via the inverse
neutron $\beta$-decay
\begin{equation}
    \bar{\nu}_e+p \to e^{+}+n\,.
    \label{eq:decay}
\end{equation} 
In Fig.~5 of Ref.~\cite{kamlandPRL} the spectral data are given in 13
bins of prompt energy above 2.6 MeV. To simulate the KamLAND data we
calculate the expected number of events in each bin for given
oscillation parameters as
\begin{equation}\begin{split}\label{eq:noe}
    N_i^\mathrm{th}(\Delta m^2,\theta) &= f
    \int dE_\nu  \, \sigma(E_\nu) \times
    \\
    \sum_j \phi_j(E_\nu) & P_j(E_\nu,\Delta m^2,\theta) 
    \int_i dE_e \, R(E_e,E'_e)\,.
\end{split}\end{equation}
Here $R(E_e, E_e')$ is the energy resolution function and $E_e, E_e'$
are the observed and the true positron energy, respectively, and we
use an energy resolution of
$7.5\%/\sqrt{E(\mathrm{MeV})}$~\cite{kamlandPRL}. The neutrino energy
is related to the positron energy by $E_\nu=E_e'+\Delta$, where
$\Delta$ is the neutron-proton mass difference. The integration
interval over $E_e$ is determined by the prompt energy interval in
each bin. The neutrino spectrum $\phi(E_\nu)$ from nuclear reactors is
well known, we are using the phenomenological parameterization given
in Refs.~\cite{Vogel:iv,Murayama:2000iq}. We adopt the average fuel
composition for the nuclear reactors given in Ref.~\cite{kamlandPRL}.
Note that possible effects due to time variations in the fuel
composition have been shown to be small~\cite{Murayama:2000iq}.  The
sum over $j$ in Eq.~\eqref{eq:noe} runs over 16 nuclear plants, taking
into account the different distances from the detector and the power
output of each reactor (see Table~3 of Ref.~\cite{kamlandproposal}).
The relevant detection cross section $\sigma(E_\nu)$ is given in
Ref.~\cite{Vogel:1999zy}. In the 2-neutrino framework the
survival probability for the neutrinos coming from the reactor
$j$ is given by
\begin{equation}
    P_j(E_\nu,\Delta m^2,\theta) = 1 - 
    \sin^22\theta \sin^2\frac{\Delta m^2 L_j}{4E_\nu} \,.
\end{equation}
The normalization factor $f$ in Eq.~\eqref{eq:noe} is determined in
such a way that for the case of no oscillations we obtain a total
number of events of 86.8, as expected from the Monte-Carlo simulation
used in Ref.~\cite{kamlandPRL}.

For the statistical analysis we use the $\chi^2$-function
\begin{equation}\label{eq:chi2}
    \chi^2 = \sum_{i,j} (N_i^\mathrm{th} - N_i^\mathrm{obs})
    S^{-1}_{ij} (N_j^\mathrm{th} - N_j^\mathrm{obs}) \,.
\end{equation}
The observed number of events $N_j^\mathrm{obs}$ in each bin can be
read off from Fig.~5 of Ref.~\cite{kamlandPRL}. In the covariance
matrix $S$ we include the experimental error in each bin $\sigma_i$
(obtained from the same figure), which we assume to be uncorrelated,
and the systematic error $\sigma_\mathrm{syst} =
0.0642$~\cite{kamlandPRL} implied by the uncertainty on the total
number of events expected for no oscillations:
\begin{equation}\label{eq:S}
    S_{ij} = \sigma_i^2 \delta_{ij} +
    \sigma^2_\mathrm{syst} N_i^\mathrm{th} N_j^\mathrm{th} \,.
\end{equation}
This $\chi^2$ definition assumes Gaussian distribution of the data. For
the discussion of an alternative analysis based on poisson distributed
data see Sec.~\ref{sec:stability}.

\begin{figure} \centering
    \includegraphics[width=0.98\linewidth]{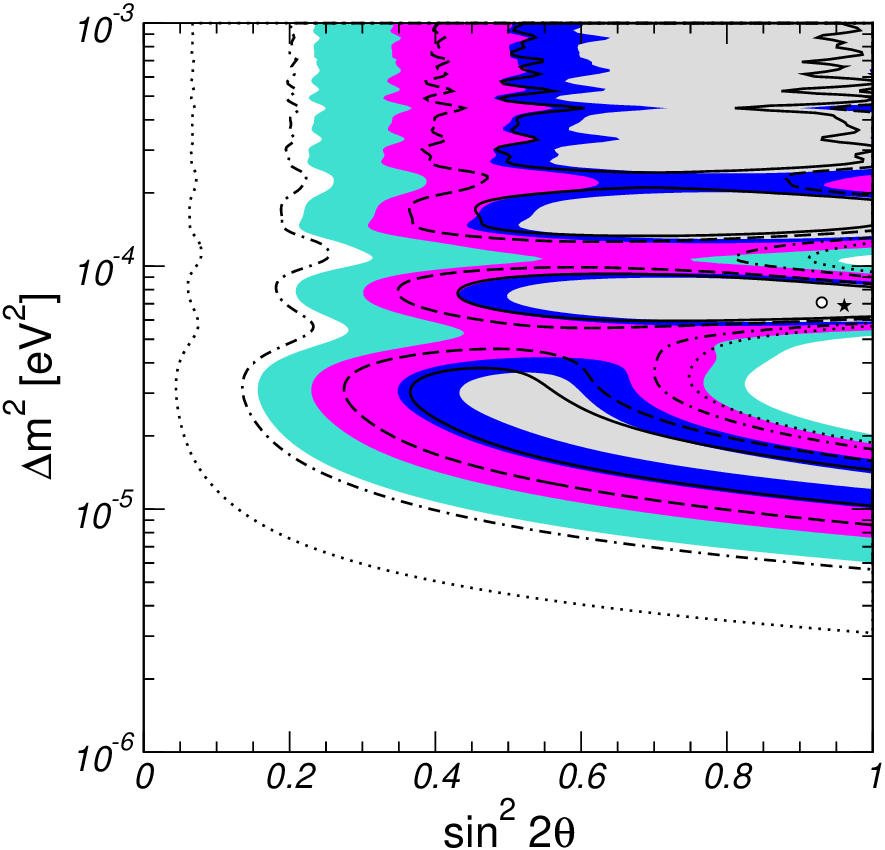}
    \caption{\label{fig:kamland} Allowed regions at 90\%, 95\%, 99\%
      and 99.73\% \CL\ (2~\dof) and best fit point from KamLAND
      spectral data.  The colored regions and the star are obtained by
      using the $\chi^2$ of Eq.~(\ref{eq:chi2}) based on the Gaussian
      approximation, the hollow regions and the dot correspond to the
      $\chi^2$ of Eq.~(\ref{eq:pois}) implied by Poisson--distributed
      data.}
\end{figure}

\section{Results and Discussion}
\label{sec:results}

\begin{figure} \centering
    \includegraphics[width=0.98\linewidth]{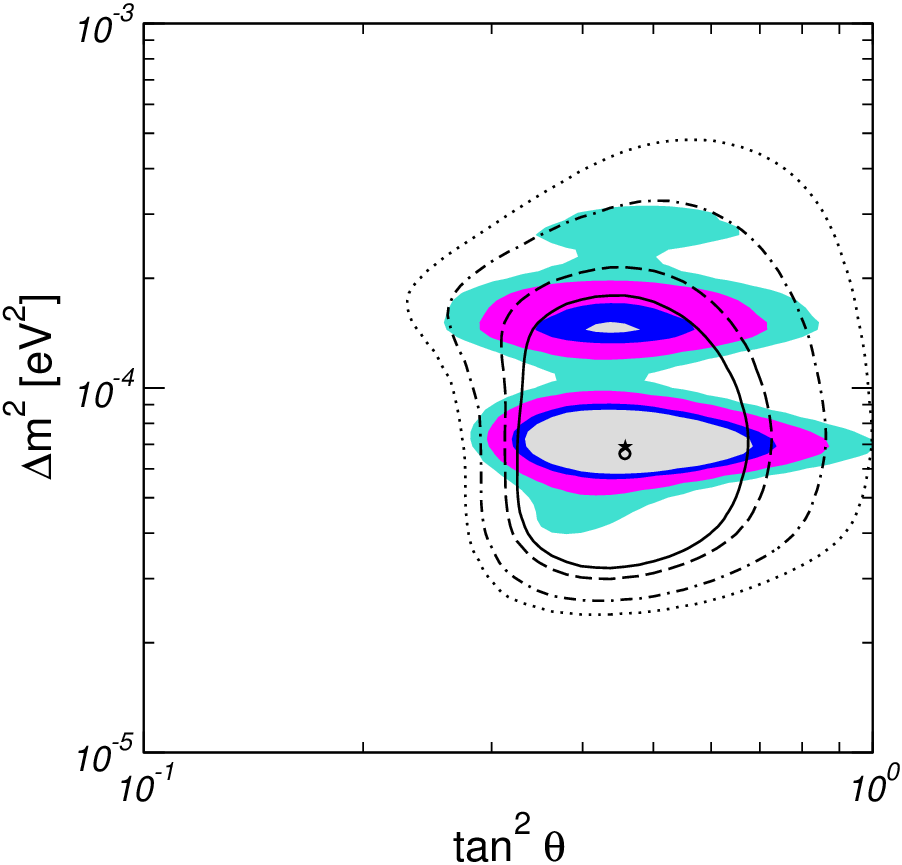}
    \caption{\label{fig:region}%
      Allowed regions at 90\%, 95\%, 99\% and 99.73\% \CL\ (2~\dof)
      from the combined analysis of solar, Chooz and KamLAND data. The
      hollow lines are the allowed regions from solar and Chooz data
      alone. The star (dot) is the best fit point from the combined
      (solar+Chooz only) analysis.}
\end{figure}

\begin{figure*} \centering
    \includegraphics[width=0.98\linewidth]{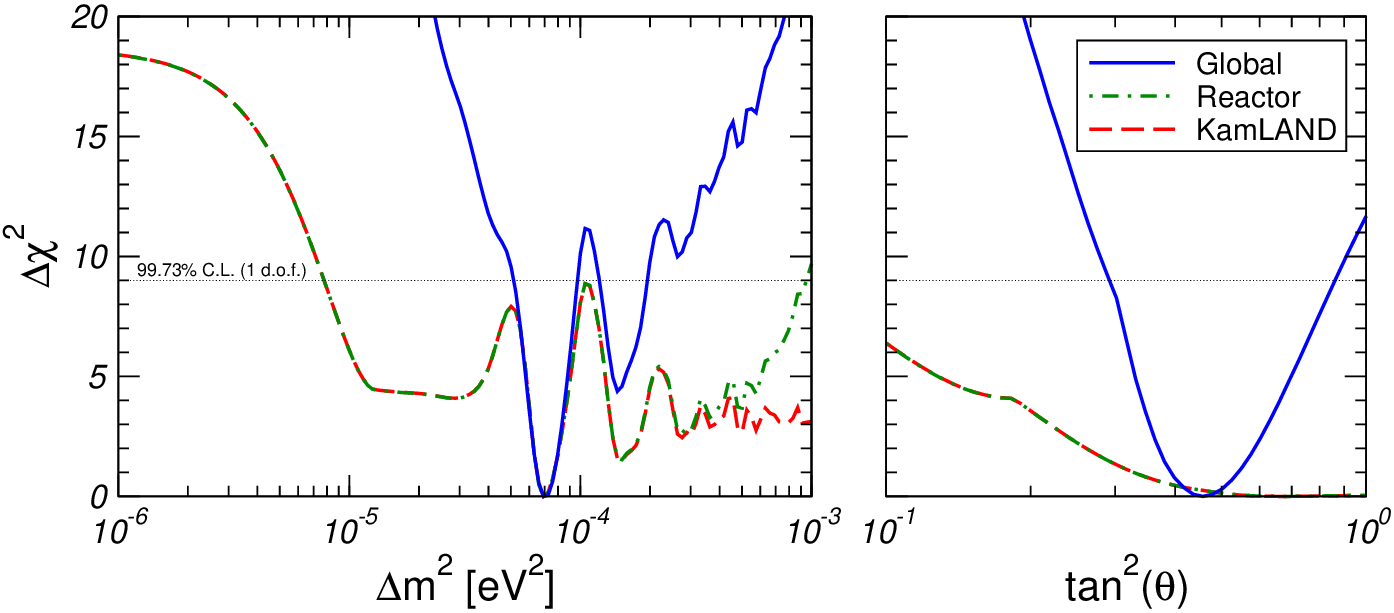}
    \caption{\label{fig:chisq}%
      \Dcq\ versus \Dms\ and $\tan^2 \theta$. The red dashed line
      refers to KamLAND alone. The green dot-dashed line corresponds
      to the full reactor data sample, including both KamLAND and
      Chooz. The blue solid line refers to the global analysis of the
      complete solar and reactor data.}
\end{figure*}

Our results are summarized in Figs.~\ref{fig:kamland}, \ref{fig:region}
and \ref{fig:chisq}.
In Fig.~\ref{fig:kamland} we show the allowed regions of the
oscillation parameters obtained from our re-analysis of the KamLAND
data. It is in good agreement with the analysis performed by the
KamLAND group, shown in Fig.~6 of Ref.~\cite{kamlandPRL}. This gives us
confidence on our simulation of the KamLAND data and therefore
encourages us to use it in a full analysis combining also with the
solar data sample.
Figs.~\ref{fig:region} and \ref{fig:chisq} show the corresponding results
obtained in a combined fit of the full KamLAND data sample with the
global sample of solar neutrino data, as well as the Chooz result. The
solar data we are using and the details of our solar neutrino analysis
are given in Ref.~\cite{Maltoni:2002ni}.

First of all, we have quantified the rejection of non-LMA solutions
and found that it is now more robust. For example, for the LOW
solution we have $\Dcq_\text{LOW--LMA} = 26.9$, which for 2~\dof\
($\Dms$ and $\theta$) corresponds to a relative probability of $1.4 \times
10^{-6}$, assuming Gaussian errors. A similar result is also found for
the VAC solution. Apart from selecting out LMA as the unique solution
of the solar neutrino problem we find, however, that the new reactor
results have little impact on the location of the best fit point:
\begin{equation}\label{eq:bfp}
    \tan^2\theta = 0.46, \qquad \Dms = 6.9\times10^{-5}~\eVq.
\end{equation}
In particular the solar neutrino mixing remains significantly
non-maximal, a point which is rather important for model-building.
Indeed bi-maximal mixing models are
disfavored~\cite{Chankowski:2000fp} while models where the solar
mixing can be non-maximal~\cite{Babu:2002dz} are preferred, as before.
This is not in contradiction with the fact that KamLAND data alone
prefer maximal mixing~\cite{kamlandPRL}, since such preference has no
statistical significance. Indeed, one can see from the right panel in
Fig.~\ref{fig:chisq} that \Dcq\ is rather flat with respect to the
mixing angle for $\tan^2 \theta \gtrsim 0.4$. This explains why the
addition of the KamLAND data has no impact whatsoever in the
determination of the solar neutrino oscillation mixing. The allowed
$3\sigma$ region we find for $\theta$ is:
\begin{equation}
    0.29 \leq \tan^2\theta \leq 0.86,
\end{equation}
practically identical to the pre-KamLAND range given in Eq.~(4) of
Ref.~\cite{Maltoni:2002ni}.

On the other hand, the new data do have a strong impact in narrowing
down the allowed range of \Dms. From the left panel of
Fig.~\ref{fig:chisq} one can read off that KamLAND data alone provides
the bound $\Dms > 8\times 10^{-6}~\eVq$, whereas the CHOOZ experiment
gives $\Dms < 10^{-3}~\eVq$, both at $3\sigma$. Hence, global reactor
neutrino data provide a robust allowed interval for \Dms, based only
on terrestrial experiments using artificial neutrino sources. However,
combining this information from reactors with the solar neutrino data
leads to a significant reduction of the allowed range: As clearly
visible in Fig.~\ref{fig:region}, the original LMA region is now split
into two sub-regions. From Fig.~\ref{fig:chisq} we obtain at $3\sigma$
(1~\dof)
\begin{gather}
    5.1\times 10^{-5}~\eVq \leq \Dms \leq 9.7\times 10^{-5}~\eVq, \\
    1.2\times 10^{-4}~\eVq \leq \Dms \leq 1.9\times 10^{-4}~\eVq. 
    \label{eq:upper}
\end{gather}
The local minimum in the region (\ref{eq:upper}) occurs for 
\begin{equation}
    \tan^2\theta = 0.42, \qquad \Dms = 1.4\times10^{-4}~\eVq
\end{equation}
with a $\Delta \chi^2 = 4.5$ with respect to the best fit point given
in Eq.~(\ref{eq:bfp}). This ambiguity might be resolved when more
KamLAND data have been collected (see
e.g.~Refs.~\cite{Murayama:2000iq,deGouvea:2001su,Barger:2000hy}).

\section{Stability of the statistical analysis}
\label{sec:stability}

The current KamLAND data sample consists of 54 anti-neutrino events,
which are distributed over the 13 energy bins.  This leads to rather
small numbers of events in each bin. The 5 bins with highest energies
contain no event at all. In such a case the use of a $\chi^2$-function
based on Poisson statistics might be appropriate. In order to check
the stability of our results we have performed also an analysis by
using \cite{pdg}
\begin{equation}\label{eq:pois}
\chi^2 = 2 \sum_i \left[ 
\alpha N_i^\mathrm{th} - N_i^\mathrm{obs} +
N_i^\mathrm{obs} \ln \left( 
\frac{N_i^\mathrm{obs}}{\alpha N_i^\mathrm{th}}
\right) \right] +
\left( \frac{1-\alpha}{\sigma_\mathrm{syst}} \right)^2
\end{equation}
where the term containing the logarithm is absent in bins with no
events. We minimize with respect to $\alpha$ in order to take into
account the overall uncertainty of the theoretical predictions.

The analysis of KamLAND data using Eq.~(\ref{eq:pois}) is shown in
Fig.~\ref{fig:kamland} as the hollow lines. We observe that this
analysis is somewhat less constraining compared to the analysis based
on the Gaussian $\chi^2$ of Eq.~(\ref{eq:chi2}). One notices that
smaller values of the mixing angle are allowed, especially at high
convidence level.  Let us note, however, that the allowed regions from
the Gaussian analysis are in better agreement with the analysis done
by the KamLAND group.  This is the reason why we prefer to use this
method for analyzing KamLAND data.  The better agreement with the
original KamLAND analysis might be related to the fact that the
inclusion of the information on the experimental errors provided by the
KamLAND collaboration in Fig.~5 of Ref.~\cite{kamlandPRL} can only be
included by means of a Gaussian $\chi^2$-function, as in
Eq.~(\ref{eq:chi2}). In this way it is possible to take into account
the asymmetric errors and the error bars in bins where the number of
events is zero.

\begin{figure} \centering
    \includegraphics[width=0.98\linewidth]{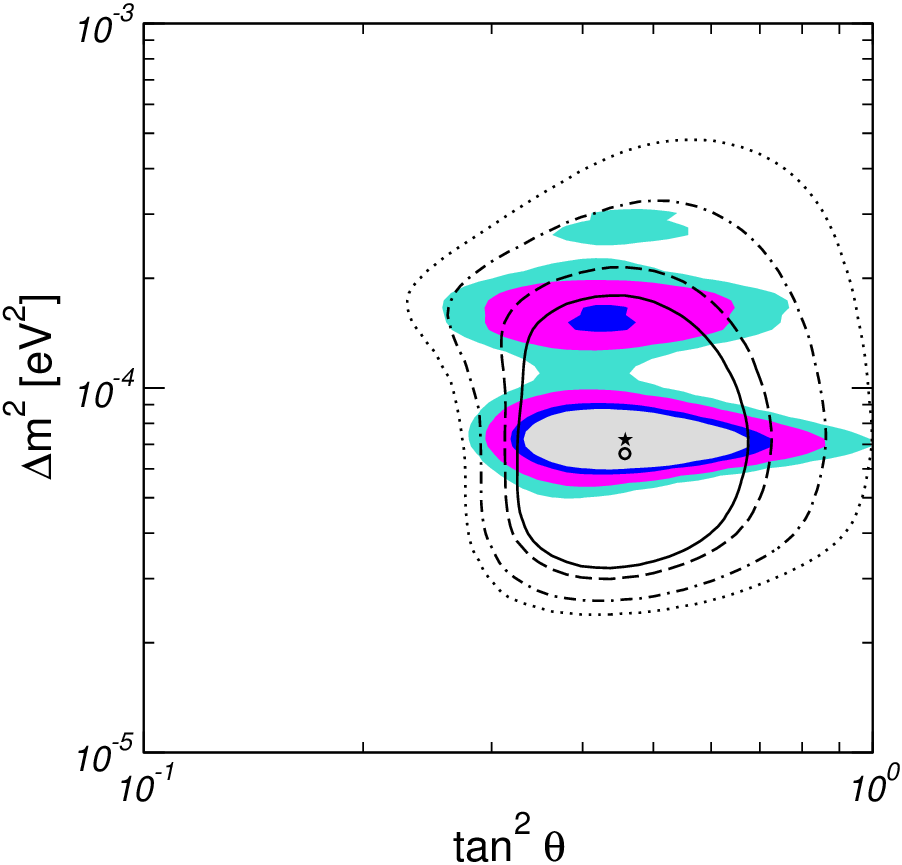}
    \caption{\label{fig:region-pois} As Fig.~\ref{fig:region}, but using for
    the analysis of KamLAND data the $\chi^2$ of Eq.~(\ref{eq:pois})
    implied by Poisson distributed data}
\end{figure}

However, we note that the determination of \Dms\ is rather stable,
only the constraint on the mixing angle is somehow affected. Since the
bound on the mixing angle in the combined analysis is dominated by
solar data, we expect the difference between the two methods to be
small after combining KamLAND with solar data. The results of this
exercise are shown in Fig.~\ref{fig:region-pois}. Comparing this
figure with Fig.~\ref{fig:region} we find indeed, that the result is
very similar.  The location of the best fit point and the 90\%, 95\%
and 99\% \CL\ regions around the best fit point are nearly identical.
However, the local minimum does not appear at the 90\% \CL, though its
location is, again, very stable. Some small differences are visible
for the 99.73\% \CL\ contour.

To summarize, although there are some notable differences between the
allowed regions obtained by assuming Gaussian or Poisson
$\chi^2$-functions for the KamLAND data taken alone, the differences
are very small when combined with solar data.  This illustrates the
robustness of our results against variations in the statistical
analysis.

\section{Conclusions}
\label{sec:conclusions}

We find that among all previous oscillation solutions to the solar
neutrino anomaly, the new reactor results from the KamLAND experiment
single out the LMA solution, rejecting all other oscillation solutions
at a significant level. Furthermore, we find that already these first
145.1 days of KamLAND data lead to a significant improvement in
narrowing down the allowed range of \Dms\ when combined with solar
neutrino data. The original LMA region now is split into two
relatively narrow islands around the values of $\Dms = 6.9\times
10^{-5}$ eV$^2$ (best fit point) and $\Dms = 1.4\times 10^{-4}$ eV$^2$
(local minimum).  However, our full analysis indicates that the new
data have little impact on the determination of the mixing angle. In
particular the solar neutrino mixing remains significantly non-maximal
($3\sigma$).

Before closing, let us note that we have considered here only the
simplest case of two neutrinos. Analyzing in detail the impact of the
KamLAND results on three-neutrino oscillation
scenarios~\cite{Gonzalez-Garcia:2000sq} and the resulting constraints
is beyond the scope of this short note.  
The improved determination of \Dms\ can also play an interesting role
in probing fine details of solar physics~\cite{Burgess:2002we}, matter
effects~\cite{Fogli:2002hb}, probing electro-magnetic neutrino
properties~\cite{Grimus:2002vb} (see also \cite{em_others}) or testing
CPT invariance in the neutrino sector~\cite{Bahcall:2002ia}.
Similarly, the nailing down of LMA as the solution has also
implications for non-oscillation solutions to the neutrino
anomaly~\cite{Valle:2002tm} in terms of spin-flavor
precession~\cite{Barranco:2002te,Friedland:2002pg}, non-standard
interactions~\cite{Guzzo:2001mi} or neutrino decay~\cite{decay}.
Clearly none can now be the leading explanation to the solar neutrino
anomaly~\cite{Pakvasa:2003zv}, although a detailed evaluation must be
performed to decide, in each case, to what extent these solutions are
now rejected.

\acknowledgments This work was supported by Spanish grant
BFM2002-00345, by the European Commission RTN grant
HPRN-CT-2000-00148, by the ESF \emph{Neutrino Astrophysics Network}
and by the \emph{Sonderforschungsbereich 375-95 f\"ur Astro-Teilchenphysik}
der Deutschen Forschungsgemeinschaft (T.S.). M.M.\ is supported by the Marie
Curie contract HPMF-CT-2000-01008.

\end{document}